\documentclass[
reprint,
twocolumn,
superscriptaddress,
showpacs,preprintnumbers,
amsmath,amssymb,
aps,prl, 
nofootinbib,
longbibliography, 
floatfix,
]{revtex4-1}

\usepackage{units}
\newcommand{\nfrac}[2]{{#1}/{#2}}

\usepackage{upgreek}
\usepackage{textcomp}
\usepackage{enumitem}  
\usepackage{graphicx}
\usepackage{epstopdf}
\usepackage{dcolumn}
\usepackage{bm}
\usepackage{hyperref}

\usepackage{xcolor}
\definecolor{mylinkcolor}{rgb}{0.0,0.0,0.66}
\hypersetup{colorlinks=true,linkcolor=mylinkcolor,urlcolor=mylinkcolor,citecolor=mylinkcolor,filecolor=mylinkcolor,breaklinks=true,linktocpage=true,linktoc=all}

\usepackage[normalem]{ulem}
\usepackage{color,soul} 
\usepackage{lipsum}  
\usepackage[export]{adjustbox}

\newcommand{\sdist}{\kern 0.20em}
\renewcommand{\eqref}[1]{Eq.\sdist(\ref{#1})}
\newcommand{\figref}[1]{Fig.\sdist\ref{#1}}

\mathchardef\mhyphen="2D

\usepackage{color}

\emergencystretch=\maxdimen
\begin{document}
	
	
	\title{Signature of Collective Plasma Effects in Beam-Driven QED Cascades}
	\author{Kenan Qu}
	\affiliation{Department of Astrophysical Sciences, Princeton University,  Princeton, New Jersey 08544, USA}  
	\author{Sebastian Meuren}
	\affiliation{Department of Astrophysical Sciences, Princeton University,  Princeton, New Jersey 08544, USA}  
	\affiliation{Stanford PULSE Institute, SLAC National Accelerator Laboratory, Menlo Park, California 94025, USA} 
	\author{Nathaniel J. Fisch}
	\affiliation{Department of Astrophysical Sciences, Princeton University,  Princeton, New Jersey 08544, USA}

	\date{\today}

	\begin{abstract}
		QED cascades play an important role in extreme astrophysical environments like magnetars. They can also be produced by passing a relativistic electron beam through an intense laser field. Signatures of collective  pair plasma effects in these QED cascades are shown to appear, in exquisite detail, through  plasma-induced frequency upshifts in the laser spectrum.
		Remarkably, these signatures can be detected even in small plasma volumes moving at relativistic speeds. Strong-field quantum and collective pair plasma effects can thus be explored with existing technology, provided that ultra-dense electron beams are colocated with multi-PW lasers.  
	\end{abstract}
	

	\maketitle
	
	
	\textit{Introduction.}---Intriguing astrophysical environments like magnetars~\cite{kaspi_magnetars_2017, cerutti_electrodynamics_2017, lin_stringent_2020, ridnaia_peculiar_2020, li_identification_2020,bochenek_fast_2020, collaboration_bright_2020, marcote_repeating_2020, amiri_second_2019, ravi_fast_2019, bannister_single_2019, cordes_fast_2019, petroff_fast_2019}, binary neutron-star mergers ~\cite{xue_magnetar-powered_2019,price_producing_2006}, and core-collapse supernovae explosions~\cite{mosta_large-scale_2015, akiyama_magnetorotational_2003} exhibit magnetic fields  substantially exceeding the QED critical field, also known as the Schwinger field \cite{schwinger_gauge_1951}. Strong-field QED cascades  fill these astrophystical objects with high density relativistic electron-positron pairs~\cite{chen_filling_2020,timokhin_maximum_2019, gueroult_determining_2019, wang_testing_2020,abbott_gw190425_2020, ligo_scientific_collaboration_and_virgo_collaboration_gw170817_2017, palenzuela_electromagnetic_2013,anderson_magnetized_2008} such that plasma effects become important.
	However, the interplay between strong-field quantum and collective plasma effects in what might be called the  {\it QED plasma regime}  remains  poorly understood~\cite{melrose_pulsar_2016, uzdensky_plasma_2014,uzdensky_extreme_2019,zhang_relativistic_2020}.

	There is thus strong motivation to elucidate the physics of QED plasmas in laboratory experiments. Even though they cannot reproduce magnetar-strength magnetic fields, the Lorentz boost of ultra-relativistic particles allows us to probe analogous conditions and produce a {beam-driven QED cascade} with exponentially growing electron, positron, and photon densities~\cite{Supp}. Such an experiment is possible only when the critical field is significantly exceeded in the boosted frame~\cite{yakimenkoPRL19} and quantum corrections to synchrotron emission and pair production become important~\cite{di_piazza_extremely_2012,sokolov_pair_2010, hu_complete_2010,thomas_strong_2012,neitz_stochasticity_2013,bulanov_electromagnetic_2013, blackburn_quantum_2014,green_transverse_2014,vranic_all-optical_2014,blackburn_scaling_2017, vranic_multi-gev_2018,magnusson_laser-particle_2019}. 
	The QED cascade then might generate pairs at a density high enough that collective plasma effects begin to play a large role. 
	
	We show that, in fact, the combination of a $\unit[3]{PW}$ laser and a \textit{dense} $\unit[30]{GeV}$ electron beam produces a quasi-neutral pair-plasma with a density that is comparable to the critical one. Such laser systems are routinely operated in several laboratories~\cite{danson_2019}. An electron beam with $\unit[10]{GeV}$ energy and $\unit[3\times{}10^{19}]{{cm}^{-3}}$ peak density represents the state of the art, available at the FACET-II facility~\cite{FACET-II}. The electron beam parameters assumed here ($\unit[30]{GeV}$ energy, $\unit[4\times 10^{20}]{{cm}^{-3}}$ density) could be achieved at SLAC with a new bunch compressor by combining the FACET-II and LCLS-Cu linac~\cite{white_2020,white_ultra_2018, yakimenkoPRL19}.

	This beam-laser collision approach has three significant advantages over the all-optical laser-laser collision approach.  \textit{First}, in producing the pair plasma, the required laser intensity ($\unit[3 \times 10^{22}]{Wcm^{-2}}$) and laser power (a few PW) are far lower than those of the all-optical approach, which requires intensities above $10^{24}\, \mathrm{W cm}^{-2}$~\cite{bell_possibility_2008,fedotov_limitations_2010, bulanov_schwinger_2010, nerush_laser_2011,elkina_qed_2011,jirka_electron_2016,grismayer_laser_2016,zhu_dense_2016,tamburini_laser-pulse-shape_2017,gonoskov_ultrabright_2017, grismayer_seeded_2017,savin_energy_2019}, only available at large $\unit[100]{PW}$-scale laser facilities~\cite{cartlidge_light_2018,OPCPA_2019,danson_petawatt_2019}.  \textit{Second}, and very importantly, not only is the QED plasma regime easier to produce, but it is easier to observe once it is produced.  Because the intensities are lower, the average gamma factor of the produced pair plasma is also much lower. This means that, at the same pair density, the plasma frequency, which signifies collective effects, is much higher.  The beam-laser approach thus solves the coupled production-observation problem. \textit{Third}, seeding the cascade with a beam instead of a gaseous or solid target 
	results in a high degree of experimental control.  
	
	In fact, the QED plasma regime is notoriously hard to observe, both in seeded laser-laser and beam-laser collisions. The plasma is moving and expanding at relativistic speeds and its volume is similar or smaller than the skin depth for realistic laser parameters. 
	Conventional detection methods, e.g., by observing plasma instabilities like the two-stream instability~\cite{Greaves_prl_1995}, the Weibel instability~\cite{Fried_1959}, or stimulated Brillouin scattering~\cite{Edwards_prl_2016}, become very difficult or even impossible with such small plasma volumes. 
	Exploring the QED plasma regime with existing technology therefore requires a new kind of diagnostic.
	
	We show here that frequency upshifts in the laser spectrum inform importantly and in exquisite detail on the interplay between strong-field quantum and collective plasma effects. They are induced by the time-varying pair plasma density, both as it forms and as it radiates.  A frequency up/downshift occurs whenever the index of refraction changes suddenly~\cite{wilks1988frequency,Kenan_slow_ionization, Shvets2017,Nishida2012, Kenan_2018_upshift,Edwards_Chirped,Bulanov2005, Peng_PRApp2021}. 
	Here,  pair production changes the particle density and thus the plasma frequency.
	In addition,  quantum synchrotron radiation reduces the electron and positron energy, and hence their effective masses, which also increases the plasma frequency.
	As we show, both analytically and numerically, detailed signatures of these effects appear in the output laser field spectrum, which are measurable at intensities as low as $\unit[10^{22}]{W cm^{-2}}$.
	
	Thus,  remarkably,  despite the small plasma volume and despite the relativistic plasma motion, signatures of the QED plasma regime might be identified experimentally with state-of-the-art technology.
	What emerges is a compelling argument for colocating laser and beam facilities to explore QED cascades in general and  the QED plasma regime in particular.

	\textit{Frequency upshift.}---When electron-positron pairs are generated in a strong laser field, their oscillation reduces the optical permittivity, thereby upshifting the laser frequency. The frequency upshift is determined by the collective plasma parameter, i.e., the plasma frequency~\cite{kruer_physics_2003} $\omega_p= \sqrt{2n_p e^2/(\gamma m_e\varepsilon_0)}$. Here, $2n_p$ is the total density of the pair particles, $e>0$ is the elementary charge, $\varepsilon_0$ is the vacuum permittivity, $m_e$ is the electron/positron rest mass, and $\gamma$ is the Lorentz factor. 
	If a small volume, counterpropagating plasma is created, it changes the instantaneous laser frequency and wavevector by~\cite{wilks1988frequency, esarey_frequency_1990, PRL_Wood_1991, Supp}, 
	\begin{align}
		\Delta\omega(x,t) &= \omega_0 \int_{t_0}^t dt' \, \left[ \frac{\partial}{\partial T} \frac{n_p(X, T)}{n_c \gamma(X, T)}\right]_{X=x-c(t-t')}^{T=t'}, \label{frequp0}\\
		\Delta k(x,t) &= \Delta\omega(x,t)/c  - \omega_p^2(x,t) /(2\omega_0 c), \label{kupshift}
	\end{align}
	where $\omega_0$ is the input laser frequency, and $n_c$ is the corresponding critical plasma density defined as $\omega_0^2 = e^2n_c/(m_e\varepsilon_0)$. After the plasma traverses through the laser, $\Delta\omega$ becomes asymptotically identical to $c\Delta k$. 
	According to Eq.~(\ref{frequp0}) the maximum frequency upshift is $\Delta\omega/ \omega_0  \approx n_p/(n_c\gamma)$. 
	Although frequency upconversion also reduces the laser intensity and changes the laser polarization~\cite{Bulanov2005, Kenan_slow_ionization}, experimental detection of these secondary effects is more challenging than measuring the laser frequency shift. 
	


	\textit{Pair plasma generation.}---An electron beam colliding with a laser pulse induces a QED cascade during the ramp-up of laser intensity as soon as the local quantum parameter $\chi_e \equiv E^*/E_s \gtrsim 1$.
	Here, $E_s = \nfrac{m_e^2c^3}{(\hbar e)} \approx \unit[1.3 \times 10^{18}]{Vm^{-1}}$ is the critical field, and $E^*=\gamma|\bm{E}_\perp + \bm{\beta}\times c\bm{B}|$ is the electric field measured in the electron rest frame; $\bm{E}$ and $\bm{B}$ are the laser electric field and magnetic field in the laboratory frame, $\bm{\beta}$ is the electron velocity normalized to the speed of light $c$, and $\gamma = (1-\bm{\beta}^2)^{-1/2}$. 
	
	Depending on the field configuration the effective expression for $\chi_e$ changes, but the dependence of, e.g., the pair production and the photon emission probability on $\chi_e$ remains universal for ultra-relativistic particles. Therefore, a laboratory experiment can provide insights relevant for extreme astrophysical plasmas, e.g., those encountered in close proximity to magnetars.
	
	During the collision the pair density continues to grow until either the beam/laser energy is depleted or the laser intensity ramps down. For an electron beam with energy $\gamma_0 m_ec^2$ and a laser with dimensionless amplitude $a_0 \equiv eE/(m_ec\omega_0)$ the quantum parameter could reach a maximum value of $\widetilde{\chi}_e \approx 2a_0\gamma_0 (\hbar\omega_0)/(m_ec^2)$. Since each particle with $\chi_e \gtrsim 1$ continues to create new pairs, the final pair density scales with $n_p \sim \widetilde{\chi}_e n_e$, although practical constraints like the finiteness of the interaction volume and the interaction time can cause deviations from this simple relation.

	\begin{figure}[t]
		\centering
		\includegraphics[width=\linewidth,valign=t]{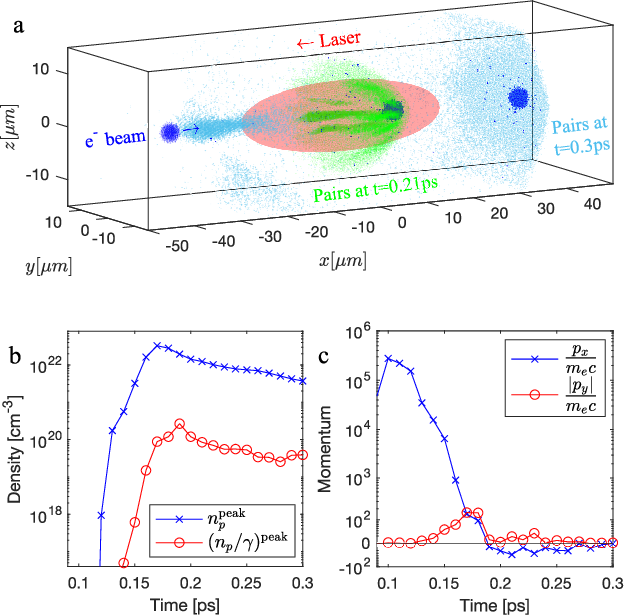}
		\caption{\textbf{a} 3D PIC simulation of a beam-driven QED cascade. An energetic, high-density electron beam (deep blue) collides with a multi-PW laser pulse (yellow), creating an electron-positron pair plasma through a QED cascade. The pair plasma is shown at $t=\unit[0.21]{ps}$ (green) and $t=\unit[0.3]{ps}$ (light blue). \textbf{b} Evolution of peak pair plasma density $n_p$ (blue), and the parameter $n_p/\gamma$ (red), which determines the laser frequency upshift. \textbf{c} Evolution of  pair particle momenta in longitudinal (blue) and transverse (red) directions, normalized to $m_ec$.} 
		\label{sch}
	\end{figure}

	\begin{figure*}[tph]
		\centering
		\includegraphics[width=\linewidth,valign=t]{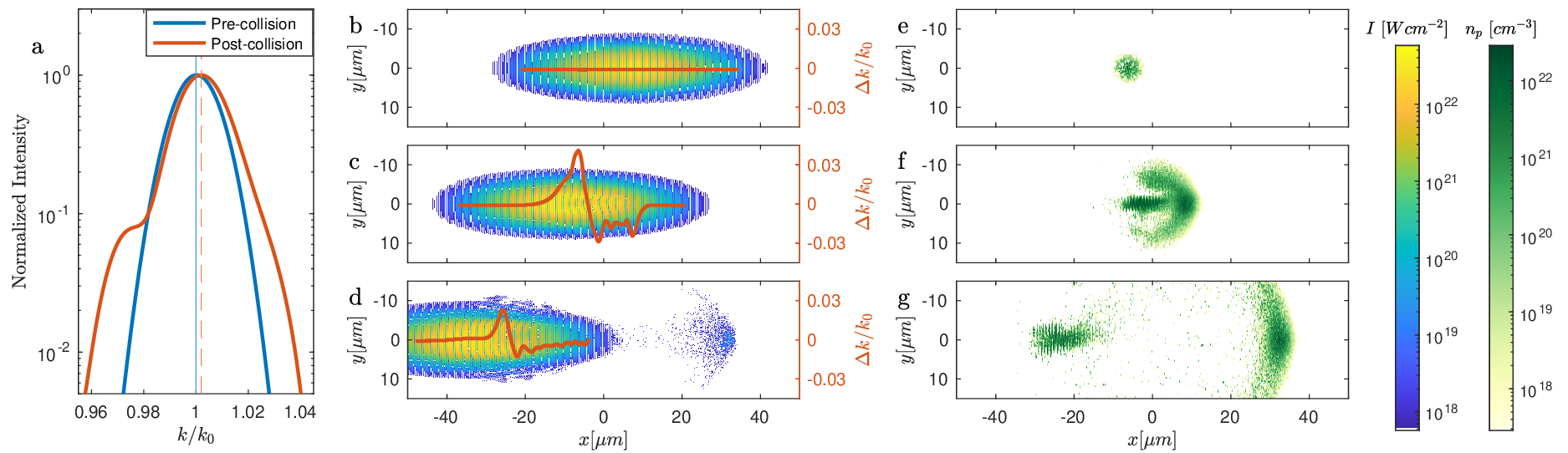}
		\caption{\textbf{a} Normalized laser intensity spectra of wavenumbers before and after the laser pulse--electron beam collision. The thin verticle lines show the peak positions. \textbf{b}-\textbf{g} Snapshots at $t=\unit[0.15]{ps}$ (\textbf{b},\textbf{e}), $t=\unit[0.2]{ps}$ (\textbf{c},\textbf{f}), and $t=\unit[0.28]{ps}$ (\textbf{d},\textbf{g}). The pseudocolor plots (\textbf{b}-\textbf{d}) show the laser beam profiles at $z=0$. The  red curve represents the instantaneous laser wavenumber through a synchrosqueezed wavelet transform of the laser field at $y=z=0$. The plots (\textbf{e}-\textbf{g}) show the pair density at $z=0$.} 
		\label{spec}
	\end{figure*}

	
	To show how a dense electron beam can indeed probe the QED plasma regime, where collective effects induce a laser frequency upshift, we carried out a ``proof-of-principle'' 3D numerical simulation with the state-of-the-art PIC code EPOCH~\cite{EPOCH2015,RIDGERS2014273}, that provides a QED module (see, e.g.,~\cite{savin_energy_2019,blackburn_scaling_2017, jirka_electron_2016, gonoskov_extended_2015}). 
	To illustrate the effect, we consider a $\unit[1]{nC}$, $\unit[300]{GeV}$ electron beam~\cite{CLIC,AWAKE}, shown as a blue sphere in Fig.~\ref{sch}(a), which collides with a counterpropagating $\unit[0.8]{\upmu{}m}$ wavelength, $24$~PW laser pulse~\cite{OPCPA_2019}, shown as an yellow spheroid~\cite{Supp}. 
	The effect is shown in detail in the 3D simulation; it can also be observed with less extreme conditions, though the exact upshifts are harder to resolve numerically.


	The electron beam sphere is injected from the left boundary ($x=\unit[-50]{\upmu m}$) with a density distribution $n_e = n_0 \exp\{-[(x+ \unit[50]{\upmu m} -ct)^2+\rho^2]/(2r_0^2)\}$, where $\rho=\sqrt{y^2+z^2}$, $n_0 = 4\times10^{20} \,\mathrm{cm}^{-3}$ and $r_0 = \unit[1]{\upmu m}$.
	The Gaussian laser pulse with linear $y$ polarization is injected from the right boundary to focus at the central plane $x=0$. The laser intensity profile is 
	$I \simeq I_0 \cdot[w_0/w(x)]^2 \exp[-2\rho^2/w^2(x)] \exp[-2(x- \unit[48]{\upmu m} +ct)^2/\tau^2]$, where $I_0 = 6\times10^{22}\, \mathrm{Wcm}^{-2}$ is the peak intensity, $w(x)=w_0\sqrt{1+(x/x_R)^2}$, $x_R=\pi w_0^2/\lambda \simeq \unit[98]{\upmu m}$ is the Rayleigh length, $w_0 = \unit[5]{\upmu m}$ is the waist, and $\tau=\unit[50]{fs}$ is the pulse duration. If the electron beam energy were not depleted by the QED cascade, these parameters would yield a quantum parameter  $\widetilde{\chi}_e \approx 220$ at the Gaussian waist in the focal plane, and $\widetilde{\chi}_e \approx 600$ at the laser focus. The simulation starts at $t=\unit[-0.205]{ps}$ and ends at $\unit[0.32]{ps}$.



	The simulation shows that the collision quickly creates a pair plasma with an exponentially growing density, as illustrated in Fig.~\ref{sch}b. The 2D cross section plots in Fig.~\ref{spec}e-g show a balloon-like plasma expansion caused by a transverse acceleration of the pairs in the strong laser field. The pairs are principally located near the core of the electron beam, and only a small amount of low-energy pairs expands in the strong laser field. The peak plasma density, shown as a blue curve in Fig.~\ref{sch}b, reaches a peak value of $n_p= 82n_0 = \unit[3.3\times 10^{22}]{cm^{-3}}$ at $t=\unit[0.17]{ps}$. At the same time, the total charge saturates at a peak value of $\unit[139]{nC}$. The laterally expanding particles move to regions with lower laser intensity and even leave the simulation box.

	\textit{Plasma deceleration.}---The parameter $n_p/\gamma$, which determines the frequency upshift, continues to grow even after the pair density $n_p$ reaches its peak value. This implies that $\gamma$ decreases faster than $n_p$ until $t=\unit[0.19]{ps}$, where $n_p/\gamma$ reaches the peak value $\unit[2.7\times10^{20}]{cm^{-3}}$.  
	
	While pair generation happens when the particle quantum parameter $\chi_e \gtrsim 1$, pairs with $\chi_e \lesssim 1$ continue to lose energy via synchrotron radiation. The energy loss remains significant as long as $\chi_e \gtrsim 0.1$. 
	Thus, in a sufficiently long laser pulse, the laser reduces the pair gamma factor asymptotically to
	\begin{gather}
		\label{eqn:gamma}
		\gamma \lesssim 0.1 \frac{\gamma_0}{\widetilde{\chi}_e} \approx \frac{0.05}{a_0} \frac{m_ec^2}{\hbar\omega_0}.
	\end{gather}
	The effect of radiation friction is shown in Fig.~\ref{sch}c, where the blue curve reveals that the pair plasma rapidly loses longitudinal momentum before $t=\unit[0.19]{ps}$.

	
	\textit{Particle reflection.}---According to the classical equations of motions the radiation pressure of a counterpropagating plane-wave laser field can (instantaneously) transfer energy of the order of $m_ec\,[a_0^2/(4\gamma)]$~\cite{di_piazza_extremely_2012}. If the symmetry of acceleration/deceleration is broken, e.g., by the emission of photons which induce a large recoil, the laser can stop and even reflect electrons/positrons~\cite{li_attosecond_2015,fedotov_radiation_2014}. As a result, \eqref{eqn:gamma} is only valid until $\gamma \sim a_0$, at which point the plasma is reflected and re-accelerated by the counterpropagating laser. Thus, we find that particle reflection is possible if
	\begin{gather} \label{eqn:reflection}
		a_0 \gtrsim \sqrt{0.05 m_ec^2/(\hbar\omega_0)},
	\end{gather}
	for a sufficiently long laser pulse.
	For optical lasers with $\hbar\omega_0 \sim \unit[1]{eV}$, the threshold is approximately $a_0 \gtrsim 100$, corresponding to $I \gtrsim 10^{22} \mhyphen 10^{23}\, \mathrm{Wcm}^{-2}$. 
	Reflection of the plasma can be observed in Fig.~\ref{sch}c and Fig.~\ref{spec}g: the longitudinal momentum becomes negative at $t=\unit[0.2]{ps}$ and the pairs are spreading throughout the simulation box at $t=\unit[0.3]{ps}$.

	
	Particle reflection 	is critically advantageous, because the maximum laser frequency upshift is induced when the plasma gamma factor reaches its minimum. Hence, we can assume $\gamma \sim a_0$ for the plasma gamma factor.  Figure~\ref{sch}c  shows that the maximum transverse momentum is $\sim a_0 m_ec$. 
	Thus, one obtains the following ``rule of thumb'' for the maximum achievable pair plasma density and the relevant gamma factor
	\begin{gather}
		\label{eqn:ruleofthumb}
		n_p \sim \widetilde{\chi}_e n_e, \quad \gamma \sim a_0,
	\end{gather}
	if the condition given in \eqref{eqn:reflection} is met and the QED cascade reaches its asymptotic state.

	\textit{Scaling laws.}---By combining \eqref{frequp0} with \eqref{eqn:ruleofthumb} one finds an order-of-magnitude estimate for the expected laser frequency upshift 
	\begin{gather}\label{frequp}
		\frac{\omega_f^2}{ \omega_0^2}  - 1 \sim \frac{\widetilde{\chi}_e n_e}{n_ca_0} \sim \gamma_0 \frac{\hbar\omega_0}{m_ec^2} \frac{n_e}{n_c}. 
	\end{gather}
	This relation is valid for an idealized model, i.e., a homogeneous electron beam counterpropagating with a plane-wave laser. It assumes that the QED cascade fully develops and that the pair plasma is eventually stopped and reflected.



	The simulation shown in Fig.~\ref{sch} yields a pair plasma which has a peak value of $n_p/\gamma$ corresponding to $6.7\%$ of the critical plasma density at rest $n_c\approx \unit[1.7 \times{}10^{21}]{cm^{-3}}$. A possible experimental setup uses two on-axis parabolic mirrors with a hole to focus and re-collect the laser. On the axis $y=z=0$, where we have the highest numerical resolution, the Fourier-transformed electric field, shown in \figref{spec}\textbf{a}, reveals an upshift of the peak wavevector by $\Delta k/k_0 \approx 0.2\%$. We also see an excess of upshifted and downshifted photons around  $\Delta k/k_0 \approx\pm 5\%$. A change of this order of magnitude is expected based on  Eq.~(\ref{kupshift}) and the peak plasma density observed in \figref{sch}\textbf{b}. The change of wavevector transforms into a change of frequency when the plasma exits the laser pulse. 
	
	A more sophisticated time-frequency analysis based on wavelet transforms~\cite{flandrin_explorations_2018, thakur_synchrosqueezing_2013, daubechies_synchrosqueezed_2011} is shown in Fig.~\ref{spec}\textbf{b}-\textbf{d}.  Such a time-frequency diagram could be measured with techniques like frequency-resolved optical gating~\cite{FROG} or spectral shear interferometry for direct electric field reconstruction~\cite{SPIDER}. The numerical analysis shows that the flattop input frequency spectrum (see Fig.~\ref{spec}b) becomes chirped at the region of plasma creation near $x=0$ in Fig.~\ref{spec}c. The chirped region propagates along the laser direction. The maximum instantaneous wavevector upshift reaches $\Delta k/k_0 \approx 2.4\%$. This amount of up/downshift is in agreement with the up/downshift observed in the front and tail of the Fourier spectrum in \figref{spec}\textbf{a}.

	\begin{figure}[th]
		\centering
		\includegraphics[width=\linewidth,valign=t]{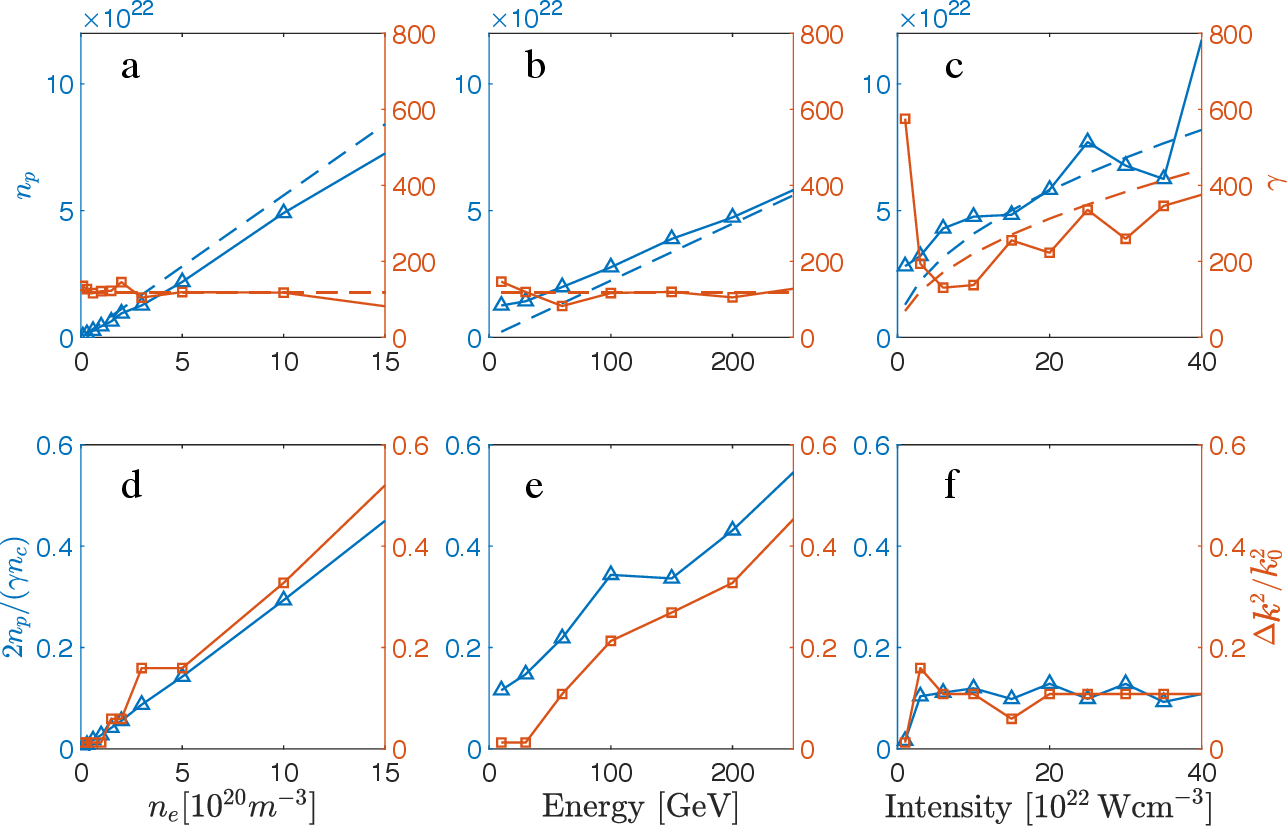}
		\caption{(a)-(c) The peak density $n_p$ and the Lorentz factor $\gamma$ of the created pair plasma. (d)-(f) Analytical predictions and numerically simulated peak values of the relative frequency upshift of the laser pulse. The numerical 1D PIC simulations (marked curves) employ: $n_e=\unit[4\times10^{19}]{{cm}^{-3}}$ peak electron beam density  [except in (a) and  (d)], $\unit[1]{\upmu m/c}$ rms duration, $\unit[100]{GeV}$ energy [except in (b) and (e)]; $\unit[3\times10^{22}]{W cm^{-2}}$ laser peak intensity [except in (d) and (f)], and $\unit[100]{fs}$ pulse duration. The dashed curves show the theoretical estimates from Eqs.~(\ref{eqn:ruleofthumb}) and~(\ref{frequp0}), respectively.} 
		\label{parascan}
	\end{figure}

	\textit{Parameter scan.}---To verify the analytical scaling laws given above, a series of 1D QED-PIC simulations were conducted with different electron beam densities, beam energies, and laser intensities. These parameter scans are possible, as 1D simulations require significantly less computational resources. They do not model transverse effects such as plasma inhomogeneity and laser diffraction, but the particle momenta and currents, which are responsible for the plasma permittivity and hence the laser frequency upshift, are effectively simulated in three dimensions.
	
	Figure~\ref{parascan}a and~b  show that either increasing the beam density or the beam energy causes a linear increase of the created pair plasma density, whereas the final gamma factor remains constant at about $\gamma \approx a_0$. Both observations are in agreement with \eqref{eqn:ruleofthumb}. The maximum frequency upshift shown in Fig.~\ref{parascan}d and~e shows a linear scaling in decent agreement with \eqref{frequp}. When the laser intensity $I$ is increased in Fig.~\ref{parascan}c and~f, both plasma density $n_p$ and gamma factor $\gamma$ increase as $\sqrt{I}$ [see \eqref{eqn:ruleofthumb}]. The parameter $n_p/\gamma$ remains constant, implying that the laser frequency upshift is independent of the laser intensity $I$ as long as the particle reflection condition [see \eqref{eqn:reflection}] is met. The reflection condition is violated for laser intensities below $3\times 10^{22} \,\mathrm{W cm}^{-2}$, causing a deviation of the frequency upshift from this scaling at very low intensities.


	According to \eqref{frequp} and Fig.~\ref{parascan}, a laser frequency upshift, reflecting collective effects, becomes  observable experimentally if the laser intensity and electron beam density are above ${\sim}\unit[10^{22}]{Wcm^{-2}}$ and ${\sim}\unit[10^{20}]{cm^{-3}}$, respectively. Such parameters require only a moderate upgrade of existing facilities, e.g., SLAC's FACET-II~\cite{FACET-II}. Indeed, a separate set of 3D QED-PIC simulations shows the tantalizing prospect: a $\unit[3]{PW}$ laser pulse ($\unit[50]{fs}$-duration, $\unit[2.5]{\upmu m}$-waist, $\unit[3 \times 10^{22}]{Wcm^{-2}}$ intensity), colliding with a $\unit[1]{nC}$, $30\, \mathrm{GeV}$, ${4\times10^{20}}\, \mathrm{cm}^{-3}$ electron beam creates an electron-positron pair plasma of $19\,\mathrm{nC}$ and peak density of $\unit[5\times{}10^{21}]{cm^{-3}}$. It causes $0.5\%$ maximum local frequency changes after the collision. While 3D-PIC simulations were not able to resolve the central frequency shift of the whole laser pulse (due to limitations of computing resources), it 
	clearly follows that an experimental measurement of the laser frequency upshift would be feasible for these parameters.

	
	Interestingly,  the produced pair plasma also  exhibits many other collective plasma effects once the parameter $n_p/\gamma$ exceeds the critical density. For example, plasma filamentation can be observed in Fig.~\ref{sch}\textbf{a} and Fig.~\ref{spec}\textbf{f}, possibly arising from the Weibel instability~\cite{Fried_1959}. 
	
	\textit{Conclusion.}---A beam-laser collision setup, together with a method of observation, solves the very challenging {joint problem} of both producing and observing the QED plasma regime. Moreover, this joint problem is solved using existing state-of-the-art beam and laser facilities, which argues compellingly for their colocation.  A key feature in this solution was to limit the pair plasma energy, thereby to increase its role in collective effects. Providing access to the QED plasma regime with available technology now offers the very real  possibility to study  in the laboratory the high energy density physics relevant to some of the very recently uncovered and most enigmatic phenomena in astrophysics.
	

	\begin{acknowledgments}
		We thank Christian Flohr Nielsen for fruitful discussions and Jean-Pierre Matte for constructive criticism that helped us to improve our work. This work was supported by NNSA Grant No. DE-NA0003871, and AFOSR Grant No. FA9550-15-1-0391. At Princeton, SM received funding from the Deutsche Forschungsgemeinschaft (DFG, German Research Foundation) under Grant No. 361969338. At Stanford, SM was supported by the U.S. Department of Energy under Award No. DE-AC02-76SF00515. 
	\end{acknowledgments}

	\bibliography{Upshift}

\end{document}